\documentclass[12pt]{iopart}
\usepackage{graphicx}
\usepackage{cite}
\usepackage{graphicx}
\usepackage{amssymb}

\usepackage[cp1251]{inputenc}
\usepackage[russian,english]{babel}
\usepackage{color}

\begin{document}

\title{Dark resonances in the field of frequency shifted
feedback laser radiation}

\author{V I Romanenko$^{1}$, A V Romanenko$^{1,2}$, L P Yatsenko$^{1}$,
G~A~Kazakov$^{3}$, A N Litvinov$^{3}$, B G Matisov$^{3}$
and Yu~V~Rozhdestvensky$^{4}$}

\address{$^{1}$ Institute of Physics, Nat. Acad. Sci. of Ukraine,
46, Nauky Ave., Kyiv 03028, Ukraine}\ead{vr@iop.kiev.ua}
\address{$^{2}$ Kyiv National Taras Shevchenko University
2, Academician Glushkov Ave, Kyiv 03022, Ukraine}
\address{$^{3}$ St. Petersburg State Polytechnical University, 29, Polytechnicheskaya st, St. Petersburg
195251, Russia}\ead{andrey.litvinov@mail.ru}
\address{$^{4}$ S. I. Vavilov State Optical Institute
12, Birzhevaya Liniya st, St. Petersburg 199034, Russia}

\begin{abstract}

We present a theory  of dark resonances in a
fluorescence of a three-level atom gas interacting with a
polychromatic field of a frequency shifted feedback (FSF) laser.
We show that conditions for the resonance observation are
optimal when the phase relations between the laser spectral
components provide generation of a light pulses train.
We study analytically the field broadening and the light
shift of the resonances.
\end{abstract}
\submitto{\jpb}

\section{Introduction}
Investigations of the coherent population trapping (CPT)
effect started at the seventies of the last
century~\cite{Ari76-333,Alz76-5,Gra78-218,Orr79-5,Alz79-209}
brought to the discovery of a number of related effects and
methods such as the electromagnetically-induced transparency
effect~\cite{Koc86,Gor89,Bol91-2593,Har97-36}, the stimulated
Raman adiabatic passage
(STIRAP)~\cite{Ore84-690,Gau88-463,Ber98-1003,Vit01}, new methods
of the laser cooling~\cite{Asp88-826}, and creation of
the miniaturized quantum frequency
standards~\cite{Kna01-1545}. The CPT effect can be used also
for the  magnetic field measurements~\cite{Nag98-31,Sch04-6409}.

In the simplest case, the CPT effect appears when a
3-level atom with two long-lived and one short-lived states
interacts with the bichromatic laser radiation, and
each spectral component of this radiation excites the
transition between the long-lived and short-lived states
(the $\Lambda$-scheme of the atom-field interaction).
In this case, there is a dip (\emph{dark resonance}) in the
fluorescence intensity  dependence on the difference frequency
between the two
spectral components when this frequency difference is
close to the transition frequency between the long-lived states.
The CPT effect is based on the atom transition   in a
``coherent dark state'', i.e., in the superposition   of
the two long-lived states which does not interact with the laser
radiation.

The dark resonance linewidth depends on the duration of the
coherent interaction of an atom with the radiation. To
increase this time, some amount of a {\em buffer gas} can be
introduced into the cell with active atoms.  The active and
buffer atom  collision  does not practically change  the
long-lived states
coherence     whereas the active atom velocity is drastically
changed. If the collision rate is  high enough, the fast
flight of the active atom through the laser beams is changed
by the slow diffusion. Therefore, the duration of coherent
atom-field interaction increases and the resonance linewidth
decreases. Moreover,  the deceleration of atomic movement along
the laser light propagation direction leads to the decreasing of a
Doppler broadening of transition between the two long-lived
states (Dicke narrowing ~\cite{Dic53}).  In a buffered cell with
rubidium active atoms, the dark resonances of 20--30 Hz
linewidth have been observed~\cite{Erh01-043813,Mer03}.

Nowadays, there is a growing interest to the dark resonances
created by the frequency comb
radiation~\cite{Alz04,Sau,Ari06-169,Vla06-609,Auz}. The dark
resonance appears when the frequency of the transition
between the long-lived states is divisible by the frequency
difference of the adjacent field components. In the
recent paper~\cite{Auz}, the dark resonances were observed in the
field of the femtosecond laser radiation. In this case, the
pulse repetition frequency sets the frequency difference
between the
nearby spectral field components.  The electromagnetically induced
transparency resonances (based on the CPT effect) were
observed in a sequence of picosecond pulses produced by a
mode-locked diode laser in~\cite{Sau}. The dark
resonances of near-100\% contrast  was observed in the field of a
free running multimode laser interacting with sodium
vapors~\cite{Alz04}. If both the frequency difference
between the two nearby frequency components and the atomic
polarization relaxation rate  caused by collisions with the buffer
gas  are much smaller than the Doppler width of the optical
transitions, the number of atoms contributing into the dark
resonance and, therefore,   the dark resonance contrast  increases
due to a large number of the polychromatic field frequency components.

The frequency shifted feedback (FSF) laser is one of the
sources of a frequency comb radiation~\cite{Yat04-183}. The
simplicity of the FSF laser radiation parameters control
makes these lasers very attractive for the dark resonances
investigation. The acousto-optic modulator (AOM) sets the
frequency difference between the two nearby spectral
components and the resonator length together with the
modulation frequency sets their phases.

The dependence of the spectral component phase on its
sequence number is quadratic in FSF lasers whereas it is
linear in   mode-locked lasers~\cite{Yat04-183}. The
spectral component phases accounting is required  in the
description of the atom-comb interaction because an atom can
interact with several spectral components simultaneously (see
figure 1).

In the present article we consider the behavior of the dark
resonances created by the field of the FSF laser radiation. We
find how the dark resonance characteristics depend  on the
AOM frequency and on the resonator length.
For the simplicity, we consider the interaction of a
three-level atom with the FSF laser field. This model was
successfully applied to the description of the interaction of real
atoms with polychromatic fields~\cite{Ari06-169}. Obtained results
are applicable to the dark resonance analysis both for cells with
a buffer gas and without~it.

In \sref{s:basic} we give an expression for the electric field of
the FSF laser and present the density matrix equation of
the three-level atom in the laser radiation field. In
\sref{s:signal} we derive the expression for the detected signal,
i.e., for the part of fluorescence which is responsible for the
dark resonance. Then in \sref{s:large} and~\ref{s:small} we
consider the limiting cases when the optical coherence
relaxation rate is large or small in comparison with the Doppler
width. Obtained results make it possible  to find the
spectral component phase relations that are optimal for
the dark resonance observation and to analyze the
light shift and the field broadening of the dark resonance. The
main results of the paper are summarized in \sref{s:conclusions}.

\section{Basic equations}
\label{s:basic}

\begin{figure}
\begin{center}\includegraphics{{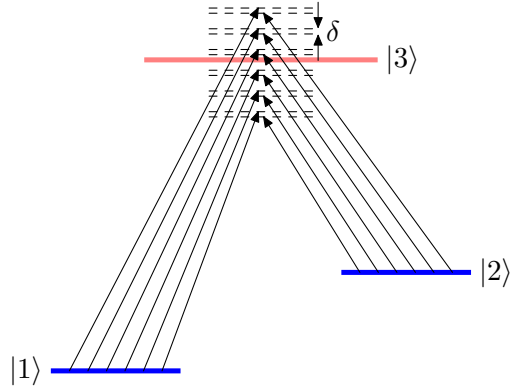}}
\caption{Scheme of interaction between the three-level atom
and the polychromatic field with equidistant spectral components
\label{fig-1}}
\end{center}
\end{figure}

Let us suppose that the laser field interacting with a three-level
atom induces the transition $|1\rangle \leftrightarrow |3\rangle$
and $|2\rangle \leftrightarrow |3\rangle$ {(see
\fref{fig-1})}. States $|1\rangle$ and $|2\rangle$ are long-lived
states (one of them can be the stable state) and the atoms can
spontaneously decay from state $|3\rangle$ to the long-lived
states with light emission. The laser field interacting with the
atoms is
\begin{equation}
{E}=\sum\limits_{n}^{}E_{n}\cos({k_{n}z-\omega_{n}t-\varphi_{n}}),
\label{eq:field-three}
\end{equation}
where $E_{n}$ and $\omega_{n}$ are the $n$-th spectral component amplitude and
frequency respectively,
$k_{n}=\omega_{n}/c$.
The frequencies of spectral components are equidistant,
\begin{equation}
\omega_{n}=\omega+\varpi{}n.
\label{eq:omega-three}
\end{equation}
Here the zero component number was chosen  in such a way
that $\omega_{0}=\omega$ corresponds to the spectral component
of maximum intensity.

We use the following model for the dependencies of $E_{n}$
and $\varphi_{n}$ on the spectral component number:
\begin{equation}
E_{n}=E_{0}\exp\left(-\frac{n^{2}}{n_{0}^{2}}\right),\qquad
\varphi_{n}=\alpha{}n^{2}+\beta{}n+\varphi_{0}. \label{eq:En}
\end{equation}
These expressions are specific for the FSF
laser~\cite{Yat04-183}.
Variation  of $\beta$  is equivalent to the change of the
time origin and therefore not affect the atomic density matrix
elements evolution. The initial phase $\varphi_{0}$ can be include
in the phase of atomic state. Hereinafter we assume
$\beta=0$, $\varphi_{0}=0$. The coefficient $\alpha$
in~(\ref{eq:En}) is determined by the  AOM frequency giving the
frequency difference $\varpi$ between adjacent spectral
components and by the round-trip time $\tau_{r}=L/c$
required for radiation to pass the resonator perimeter $L$,
\begin{equation}
\alpha=\frac{1}{2}\tau_{r}\varpi.
\label{eq:alpha}
\end{equation}

The equations for the atomic density matrix elements
$\rho_{nm}$ ($n,m=1\ldots3$) are
\begin{equation}
\eqalign{
\frac{\rmd}{\rmd t}\rho_{nm}=&-\frac{\rmi}{\hbar}\sum\limits_{l}\left(H_{nl}\rho_{lm}-\rho_{nl}{}H_{lm}\right)\\
&+\sum\limits_{kl}^{}\Gamma{}_{nm,kl}\rho_{kl}+\left(\hat{S}\rho\right)_{nm},}
\label{eq:H}
\end{equation}
where $\frac{\rmd}{\rmd t}=\frac{\partial
}{\partial{}t}+v\frac{\partial }{\partial{}z}$~is the total time
derivative, $v$~is the atomic velocity projection on the laser
radiation propagation direction, $H$ is the Hamiltonian of
the atom in the laser field, $\Gamma{}_{nm,kl}$ are the
relaxation matrix elements, $\hat{S}$ is the collisional
integral.

Let us suppose that the laser field which acts on the atom is
small so that the populations $\rho_{ii}, \, i=1,2,3$, are near to
the equilibrium. Then we can solve the set of equations for the
density matrix only for the non-diagonal elements supposing that
the diagonal elements are in equilibrium. We use the strong
collision approximation~\cite{Rau66-1176,Rau66-209} for the
collisional term describing the coherence preserving collisions
with rate $\nu$ in the equation for $\rho_{12}$. It means that
there is no any correlation between the velocity of the atom
before and after one collision. Therefore, the equations for the
non-diagonal elements of the density matrix read
\begin{eqnarray}
\frac{\rmd}{\rmd{}t}\rho_{12}=&-\rmi\sum_{n}\left(\Omega_{1,n}\rho_{32}
-\Omega_{2,n}\rho_{13}\right)\nonumber\\
&\times\cos({k_{n}z-\omega_{n}t-\varphi_{n}})\nonumber\\
&+\frac{\rmi}{\hbar}\omega_{21}\rho_{12}-\gamma_{coh}\rho_{12}\nonumber\\
&\mbox{}-{\nu}\left[\rho_{12}-M(v)N_{12}\right]
\label{eq:rho-i-ii}
,\\
\frac{\rmd}{\rmd{}t}\rho_{13}=&-\rmi\sum_{n}\left(\Omega_{1,n}\left[\rho_{33}-\rho_{11} \right]
-\Omega_{2,n}\rho_{12}\right)\nonumber\\
&\times\cos({k_{n}z-\omega_{n}t-\varphi_{n}})\nonumber\\
&+\frac{\rmi}{\hbar}\omega_{31}\rho_{13}-\gamma'\rho_{13},
\label{eq:rho-i-iii}
\\
\frac{\rmd}{\rmd{}t}\rho_{32}=&\rmi\sum_{n}
\left(\Omega_{2,n}\left[\rho_{33}-\rho_{22} \right]
-\Omega_{1,n}\rho_{12}\right)\nonumber\\
&\times\cos({k_{n}z-\omega_{n}t-\varphi_{n}})\nonumber\\
&-\frac{\rmi}{\hbar}\omega_{32}\rho_{32}-\gamma'\rho_{32},
\label{eq:rho-ii-iii}
\end{eqnarray}
where $\gamma'=\gamma_{col}+\frac{1}{2}\gamma_{sp}+\frac{1}{2}\gamma_{L}$ is
the $\rho_{13}$ and $\rho_{23}$ matrix elements relaxation rate which depends
additively on the collision rate of active atoms with buffer atoms (term $\gamma_{col}$),
on the excited state spontaneous radiation rate
$\gamma_{sp}$ and on the laser field frequency fluctuation (term with $\gamma_{L}$);
the constant $\gamma_{coh}$  describes the relaxation of the coherence of the
long-lived states (density matrix element  $\rho_{12}$);
\begin{equation}
M(v)=\frac{1}{v_{0}\sqrt{\pi}}\exp\left(-\frac{v^2}{v_{0}^{2}}\right)
\label{eq:M}
\end{equation}
is the Maxwell distribution, $v_{0}$ is the most probable velocity
of the active atom, $\hbar\omega_{nm}$ ($n,m=1,2,3$) is the
difference of the atom energies in $n$ and $m$ states,
\begin{equation}
N_{12}=\int\rho_{12}dv.
\label{eq:N}
\end{equation}
The Rabi frequencies
\begin{equation}
\Omega_{1,n}=-\frac{1}{\hbar}E_{n}d_{13},\qquad{}
\Omega_{2,n}=-\frac{1}{\hbar}E_{n}d_{23},
\label{eq:Rabi-three}
\end{equation}
characterize the interaction of $n$-th spectral field component
with the atom. We consider the Rabi frequencies as real
because we can assume without loss of generality  that the
dipole moment matrix elements $d_{13}$ and $d_{23}$ of the
transitions between the states
$|1\rangle\leftrightarrow|3\rangle$ and
$|2\rangle\leftrightarrow|3\rangle$ are real.

The normalization condition for the density matrix has the form
\begin{equation}
\sum\limits_{j=1}^{3}\int\limits_{}^{}\rho_{jj}dv=1.
\label{eq:norm}
\end{equation}

We consider small laser fields $\Omega_{10}\ll\gamma'$,
$\Omega_{20}\ll\gamma'$, so that the diagonal density matrix
elements  are close to the equilibrium values. Then we put
$\rho_{jj}=p_{j}M(v)$ into equations ~(\ref{eq:rho-i-iii}),
(\ref{eq:rho-ii-iii}) where $p_{j}$ is a probability for the
atom to be in  the state $|j\rangle$. Since the population
of the excited state is very small ($p_{3}\ll{}p_{1},p_{2}$) we
can neglect $p_{3}$ in comparison with $p_{1}$ and $p_{2}$.

Let us introduce the frequency detunings $\delta_{1}$ and
$\delta_{2}$ of the maximum intensity spectral component of the
laser field ($n=0$ in equation \eref{eq:omega-three}) from the
frequencies $\omega_{31}$ and $\omega_{32}$  of transitions
$|1\rangle \leftrightarrow |3\rangle$ and
$|2\rangle\leftrightarrow |3\rangle$ as
\begin{equation}
\eqalign{
\delta_{1}=\omega_{31}-\omega,\\
\delta_{2}=\omega_{32}-\omega=\delta_{1}-\omega_{21}}
\label{eq:Ntild}
\end{equation}
and substitute
\begin{equation}
\eqalign{
\rho_{13}=\sigma_{13}\exp\left(-\rmi{}kz+\rmi{}\omega{}t\right),\\
\rho_{32}=\sigma_{32}\exp\left(\rmi{}kz-\rmi{}\omega{}t\right).}
\label{eq:RWA}
\end{equation}
into \eref{eq:rho-i-ii}--\eref{eq:rho-ii-iii}, where $k=\omega/c$.
In rotating wave approximation  we neglect the rapidly
oscillating terms $\rme^{\pm2\rmi\omega t}$. Then
the equations (\ref{eq:rho-i-ii})--(\ref{eq:rho-ii-iii}) read
\begin{eqnarray}
\frac{\rmd}{\rmd{}t}\rho_{12}&
=-\frac{\rmi}{2}\sum_{n}\left(\Omega_{1,n}\sigma_{32}
\rme^{\rmi\Phi_{n}}
-\Omega_{2,n}\sigma_{13}
\rme^{-\rmi\Phi_{n}}
\right)\nonumber\\
&+\rmi\omega_{21}\rho_{12}-\gamma_{coh}\rho_{12}-
{\nu}\left[\rho_{12}-M(v){N}_{12}\right]
\label{eq:RWA-i-ii}
,\\
\frac{\rmd}{\rmd{}t}\sigma_{13}&
=\frac{\rmi}{2}\sum_{n}\left[\Omega_{1,n}p_{1}M(v)+\Omega_{2,n}\rho_{12}\right]
\rme^{\rmi\Phi_{n}}\nonumber\\
&+\rmi\left(\delta_{1}+kv\right)\sigma_{13}-
\gamma'\sigma_{13}
\label{eq:RWA-i-iii}
,\\
\frac{\rmd}{\rmd{}t}\sigma_{32}&
=-\frac{\rmi}{2}\sum_{n}
\left[\Omega_{2,n}p_{2}M(v)+\Omega_{1,n}\rho_{12}\right]
\rme^{-\rmi\Phi_{n}}\nonumber\\
&-\rmi\left(\delta_{2}+kv\right)\sigma_{32}-
\gamma'\sigma_{32},
\label{eq:RWA-iii-ii}
\end{eqnarray}
where $\Phi_{n}=n{\varpi}t+\varphi_{n}-n{\kappa}z$,
$\kappa={\varpi}/{c}$.
We seek for stationary solution of equations
(\ref{eq:RWA-i-ii})--(\ref{eq:RWA-iii-ii}) in the form
\begin{equation}
\eqalign{
\rho_{12}&=\sum\limits_{n}^{}r_{12,n}\rme^{\rmi\Phi_{n}},\\
\sigma_{13}&=\sum\limits_{n}^{}r_{13,n}\rme^{\rmi\Phi_{n}},\\
\sigma_{32}&=\sum\limits_{n}^{}r_{32,n}\rme^{-\rmi\Phi_{n}},
}
\label{eq:Furier}
\end{equation}
where $r_{lm,n}$ does not depend on time and coordinate.

From the equations (\ref{eq:RWA-i-iii})
and~(\ref{eq:RWA-iii-ii}), we obtain:
\begin{eqnarray}
r_{13,n}&=\frac{\rmi\Omega_{1,n}p_{1}M(v)}{2\left[\gamma'
-\rmi\left(\delta_{1}-n\varpi+n\kappa{}v+kv\right)\right]}\nonumber\\
&+\sum_{m}\frac{\rmi\Omega_{2,n-m}\rme^{\rmi\varphi_{m}-\rmi\varphi_{n}+\rmi\varphi_{n-m}}r_{12,m}}{2\left[\gamma'
-\rmi\left(\delta_{1}-n\varpi+n\kappa{}v+kv\right)\right]},
\label{eq:ri-iii}\\
r_{32,n}&=-\frac{\rmi\Omega_{2,n}p_{2}M(v)}{2\left[\gamma'+
\rmi\left(\delta_{2}-n\varpi+n\kappa{}v+kv\right)\right]}\nonumber\\
&-\sum_{m}\frac{\rmi\Omega_{1,m+n}\rme^{\rmi\varphi_{m}
+\rmi\varphi_{n}-\rmi\varphi_{m+n}}r_{12,m}}{2
\left[\gamma'+\rmi\left(\delta_{2}-n\varpi+n\kappa{}v+kv\right)\right]}.
\label{eq:riii-ii}
\end{eqnarray}
Substituting the expressions for $\sigma_{13,n}$ and
$\sigma_{32,n}$ from equation~\eref{eq:Furier} into
equations~(\ref{eq:RWA-i-ii})--(\ref{eq:RWA-iii-ii}) and taking
into account equations~(\ref{eq:ri-iii}), (\ref{eq:riii-ii}), we
obtain the equation for $r_{12,n}$ which reads
\begin{equation}
\eqalign{
&-\frac{M(v)}{4}\sum_{m}
\left(
\frac{p_{1}\Omega_{2,m}\Omega_{1,m+n}
\rme^{-\rmi\varphi_{m}-\rmi\varphi_{n}+\rmi\varphi_{m+n}}}{\gamma'-\rmi[\delta_{1}-(m+n)\varpi+kv]}\right.\\
&\left.\mbox{}+\frac{p_{2}\Omega_{1,m}\Omega_{2,m-n}
\rme^{\rmi\varphi_{m}-\rmi\varphi_{n}-\rmi\varphi_{m-n}}}{\gamma'+\rmi[\delta_{2}-(m-n)\varpi+kv]}
\right)\\
&\mbox{}+\rmi\left(\omega_{21}-n\varpi\right)r_{12,n}-\gamma_{coh}r_{12,n}\\
&\mbox{}-{\nu}\left[r_{12,n}-M(v){R}_{12,n}\right]\\
&-\frac{1}{4}\sum\limits_{m,l,j}\left(\frac{\Omega_{1,n+l}\Omega_{1,j+l}
\rme^{\rmi\varphi_{n+l}-\rmi\varphi_{n}+\rmi\varphi_{j}-\rmi\varphi_{j+l}}}{\gamma'+\rmi{}(\delta_{2}-l\varpi+kv)}\right.\\
&\mbox{}\left.+\frac{\Omega_{2,l-n}\Omega_{2,l-j}
\rme^{-\rmi\varphi_{l-n}-\rmi\varphi_{n}+\rmi\varphi_{j}+\rmi\varphi_{l-j}}}{\gamma'+\rmi{}(\delta_{1}-l\varpi+kv)}\right)r_{12,j}=0.
} \label{eq:ri-ii-comm-alt-g}
\end{equation}
Here we neglected the terms with $\kappa$ in denominators owing to
the condition $\kappa{}v/\varpi=v/c\ll1$ and
introduced
\begin{equation}
R_{12,n}=\int{}r_{12,n}dv.
\label{eq:R}
\end{equation}
We also neglected   the coordinate derivative calculating
 $\frac{\rmd}{\rmd{}t}\rho_{12}$. Therefore, we neglect the
Doppler broadening of the two-photon transition $|1\rangle
\leftrightarrow |2\rangle$. It means that we assume
$\frac{\omega_{21}}{c}v_{0}<\gamma_{col}$.

If the solution of equation \eref{eq:ri-ii-comm-alt-g} is known,
it is possible to obtain the optical coherences and the atomic
fluorescence intensity. In the next section we obtain the
expression for the fluorescence intensity describing the
dark resonance.

\section{Registered signal}
\label{s:signal}

Observing the dark resonance in the polychromatic laser
field, one registers the dependence of fluorescence intensity on
the frequency difference $\varpi$ between adjacent spectral
components. The fluorescence intensity  is proportional to
the number $S$ (hereinafter called  the signal) of
atoms excited by the laser field into the state $|3\rangle$
per unit of time,
\begin{equation}
\eqalign{
S&=\mathop{}i\sum_{n}\int\limits_{}^{}\left[\Omega_{1,n}\left(\rho_{31}-
\rho_{13}\right)+\Omega_{2,n}\left(\rho_{32}-\rho_{23}
\right)\right]\\
&\times\cos({k_{n}z-\omega_{n}t-\varphi_{n}})\,\rmd{}v.
}
\label{eq:signal}
\end{equation}

Averaging the expression~(\ref{eq:signal}) over time, we obtain
the expression for the registered signal,
\begin{equation}
\langle{S}\rangle=\sum_{n}\int\limits_{}^{}
\mathop{\mathrm{Im}}\left(
r_{13,n}\Omega_{1,n}-
r_{32,n}\Omega_{2,n}\right)\rmd{}v.
\label{eq:signal-avt}
\end{equation}
Note that this result does not depend on the coordinate.

When we substitute equations \eref{eq:ri-iii}
and~\eref{eq:riii-ii} into equation~\eref{eq:signal-avt}, the
contribution of the   summands which does not contain sums
is proportional  to the square of the laser field
amplitude. This contribution (determined by $\gamma'$ and Doppler
width $kv_0$ of the optical transition) is practically
constant in the vicinity of the CPT resonance. A  part
$\tilde{S}$ of the registered signal determined by the
CPT resonance can be obtained by the substitution of
the terms of equations~(\ref{eq:ri-iii}) and~(\ref{eq:riii-ii})
containing $r_{12,m}$ into equation~(\ref{eq:signal-avt}),
\begin{equation}
\eqalign{
\tilde{S}& = \mathop{\mathrm{Re}}\int\limits_{}^{}\sum_{m,n}\left(
\frac{\Omega_{1,n}\Omega_{2,n-m}\rme^{\rmi\varphi_{m}-\rmi\varphi_{n}+\rmi\varphi_{n-m}}}{2\left[\gamma'-\rmi(\delta_{1}-n\varpi+kv)\right]}\right.\\
&\left.\mbox{}+\frac{\Omega_{1,n}\Omega_{2,n-m}\rme^{\rmi\varphi_{m}-\rmi\varphi_{n}+\rmi\varphi_{n-m}}}{2\left[\gamma'+\rmi(\delta_{2}-(n-m)\varpi+kv)\right]}\right)r_{12,m}\,\rmd{}v.
}
\label{eq:Sw}
\end{equation}

In the next sections we consider the signal $\tilde{S}$ for
different relations between the optical coherence relaxation rate
$\gamma'$, the Doppler width $kv_{0}$, the frequency
$\omega_{21}$ of the transition between the two long-lived
states, and  the width $n_{0}\varpi$ of the FSF laser
radiation spectrum.

\section{Large optical coherence relaxation rate}
\label{s:large} When  the optical coherence relaxation rate
is larger than the Doppler width
\begin{equation}
\gamma'\gg{}kv_{0},\quad
\label{eq:cr-kv}
\end{equation}
we can neglect the velocity containing terms in denominators of
\eref{eq:ri-ii-comm-alt-g} and \eref{eq:Sw} and perform the
integration over velocity. Integration of the signal~(\ref{eq:Sw})
gives the expression for the signal that contains $R_{12,m}$. The
equations for $R_{12,m}$ can be found by integration of equation
\eref{eq:ri-ii-comm-alt-g} over velocity. It is important that
after this integration, equation \eref{eq:ri-ii-comm-alt-g} does
not depend on the coherence preserving collision rate $\nu$.
So, we conclude that the amplitude and the shape of the dark
resonance do not depend on $\nu$ if inequality \eref{eq:cr-kv} is
fulfilled. Below we consider the cases of ``broad-band'' and
``narrow-band'' (in comparison with $\gamma'$) spectrum of laser
radiation.

\subsection{The narrow-band spectrum of laser radiation}
In this subsection we consider the case when the
expression~(\ref{eq:Sw}) for the signal takes an especially
simple form. Let us suppose that the optical coherence
relaxation rate $\gamma'$ is large in comparison with the laser
spectrum width $n_{0}\varpi$ and with the detuning $\delta_{1}$  
of the laser spectral component of maximal intensity
from the transition frequency between states $|1\rangle$ and
$ |3\rangle$,
\begin{equation}
\gamma'\gg{}n_{0}\varpi,\quad
\gamma'\gg{}|\delta_{1}|.\label{eq:crit-kv}
\end{equation}%
As a result of the first inequality in~(\ref{eq:crit-kv}),
an inequality $\gamma'\gg{}\omega_{21}$ is also valid.

\begin{figure}
\begin{center}
\includegraphics{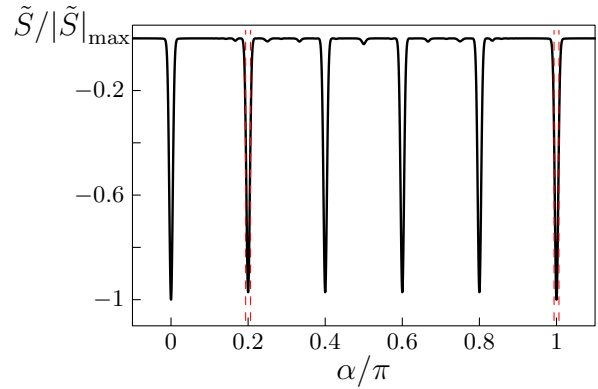}
\caption{The normalized signal at two-photon resonance
($\delta=0$) versus $\alpha$ for $n_{0}=10$, $\tilde{m}=5$, $\varpi=10\gamma_{coh}$.
Dashed line corresponds to
the difference $\Delta\alpha$ between $\alpha$ and $\alpha_{1}$ \label{fig-2}}
\end{center}
\end{figure}

It is evident from the structure of equations
\eref{eq:ri-ii-comm-alt-g} that all $r_{12,n}$ are small in
comparison with $r_{12,\tilde{m}}$, where $\tilde{m}$ is
determined from the condition  of
$\left|\omega_{21}-\tilde{m}\varpi\right|$ minimum. Taking into
account only the summand with $j=\tilde{m}$ in equation for
$r_{12,\tilde{m}}$ we see that if the conditions \eref{eq:cr-kv}
and \eref{eq:crit-kv} are fulfilled, the coefficient at
$r_{12,\tilde{m}}$ in the second sum of equation
\eref{eq:ri-ii-comm-alt-g} is real. Therefore, the second
sum in equation \eref{eq:ri-ii-comm-alt-g} describes the field
broadening. This broadening is small in comparison with
$\gamma_{coh}$ if
\begin{equation}
\gamma'\gamma_{coh}\gg{}n_{0}(\Omega_{1,0}^{2}+\Omega_{2,0}^{2}).
\label{eq:cr-broad}
\end{equation}
In this subsection we assume that the inequality
\eref{eq:cr-broad} is valid and, therefore, we neglect the
field broadening. Then, from the
equations~\eref{eq:ri-ii-comm-alt-g} and \eref{eq:Sw}, we obtain
\begin{equation}
\eqalign{
\tilde{S}&=-\mathop{\mathrm{Re}}\sum_{n,m,l}
\frac{\Omega_{1,n}\Omega_{2,n-{m}}\Omega_{1,l}\Omega_{2,l-{m}}
}{4\gamma'^{2}
\left[\gamma_{coh}-\rmi(\delta+(\tilde{m}-m)\varpi)\right]}\\
&\times{}\rme^{-\rmi\varphi_{n}+\rmi\varphi_{n-{m}}+\rmi\varphi_{l}-\rmi\varphi_{l-{m}}},
}
\label{eq:S_large}
\end{equation}
where the two-photon detuning
\begin{equation}
\delta=\omega_{21}-\tilde{m}\varpi
\label{eq:d}
\end{equation}
is introduced. The main contribution into the sum over $m$ in
the expression~(\ref{eq:S_large}) comes from summands with
$m=\tilde{m}$ because the  frequency difference $\varpi$ between
adjacent spectral components is much larger than the relaxation
rate $\gamma_{coh}$ of the coherence between $|1\rangle$ and
$|2\rangle$ states. An example of the dependence of the signal
(\ref{eq:S_large}) at two-photon resonance ($\delta=0$) on
the parameter $\alpha$ determining the spectral field
component phases is presented in \fref{fig-2}. One can see that
the signal is maximal when $\alpha$ is divisible by
$\pi/\tilde{m}$. These values of $\alpha$ can be also obtained
from the expression for the linear combination of the phases
in the exponent in the expression~(\ref{eq:S_large})
\begin{equation}
\varphi_{l}-\varphi_{n}+\varphi_{n-m}-\varphi_{l-m}=2\alpha{}m(l-n).
\label{eq:phi}
\end{equation}
The right hand side of the equation~(\ref{eq:phi}) is
divisible by $2\pi$ for the summands with $m=\tilde{m}$ (which
give the main contribution into the signal) and with any $l$
and $n$, if $\alpha=\alpha_{j}$ where
\begin{equation}
\alpha_{j}=\frac{\pi}{\tilde{m}}j,\qquad\mbox{$j$ is integer.}
\label{eq:a-opt}
\end{equation}
Therefore, the contributions of the terms with different $n$
and $l$ are summed. If the difference between $\alpha$ and
$\alpha_{j}$ is of order of
\begin{equation}
\Delta\alpha=1/(mn_{0}),
\label{eq:dalpha}
\end{equation}
the part of summands in expression~(\ref{eq:S_large}) has different
signs, and the signal magnitude decreases.
\Fref{fig-2} shows that
the equation~(\ref{eq:dalpha}) is in good accordance with the
numerical simulation result: if $|\alpha-\alpha_{j}| \gtrsim
2\Delta\alpha$ the signal practically vanishes.

This effect can be explained qualitatively in a different
way. The scheme of the interaction between an atom and the laser
field shown in \fref{fig-1} can be considered as a set of
$\Lambda$-systems. We will call the $\Lambda$-system formed
by $n$-th and $p$-th laser field component acting at
$|1\rangle\leftrightarrow|3\rangle$ and
$|2\rangle\leftrightarrow3\rangle$ transition respectively as
$\Lambda_{n,p}$. For each  $\Lambda_{n,p}$-system, the dark
state, i.e. the coherent superposition of states $|1\rangle$ and
$|2\rangle$ that does not interact with $n$-th and $p$-th
components of laser radiation, exists. However, these dark
states depend on the relative phase
$\varphi_{n,p}=\varphi_{n}-\varphi_{p}$ of laser components acting
on the arms of $\Lambda_{n,p}$-system. From the
equation~(\ref{eq:En}) we can see that $\varphi_{n,n-m}=2\alpha m
n-\alpha m^2$. 
If the frequency difference $\varpi$ between 
adjacent laser components becomes larger than the coherence relaxation 
$\gamma_{coh}$, i.e. $\varpi\gg \gamma_{coh}$, only
$\Lambda_{n,n-\tilde{m}}$-systems with an arbitrary $n$ gives the
contribution into the dark resonance. The dark states of
$\Lambda_{n,n-\tilde{m}}$- and $\Lambda_{l,l-\tilde{m}}$-systems
coincide when
$\varphi_{n,n-\tilde{m}}-\varphi_{l,l-\tilde{m}}=2\alpha \tilde{m}
(n-l)$ is divisible by $2\pi$ and therefore interference between
them is constructive. This condition is fulfilled for any numbers
of $n$ and $l$ only when $\alpha=\alpha_{j}$. 
If the condition $|\alpha-\alpha_j|>\Delta\alpha$ 
is fulfilled for any $j$, the destructive interference between 
various $\Lambda$-systems 
(the ``dark state'' for one $\Lambda$-system occurs to be the most light 
absorbing state for another $\Lambda$-system, etc.) 
leads to the considerable decreasing of the dark resonance amplitude.
It is the interference of the different $\Lambda$-schemes
contribution to the CPT resonance signal of an atom in a
polychromatic field that distinguish behavior of the
CPT resonances in the polychromatic and bichromatic fields.

\begin{figure}
\begin{center}
\includegraphics{{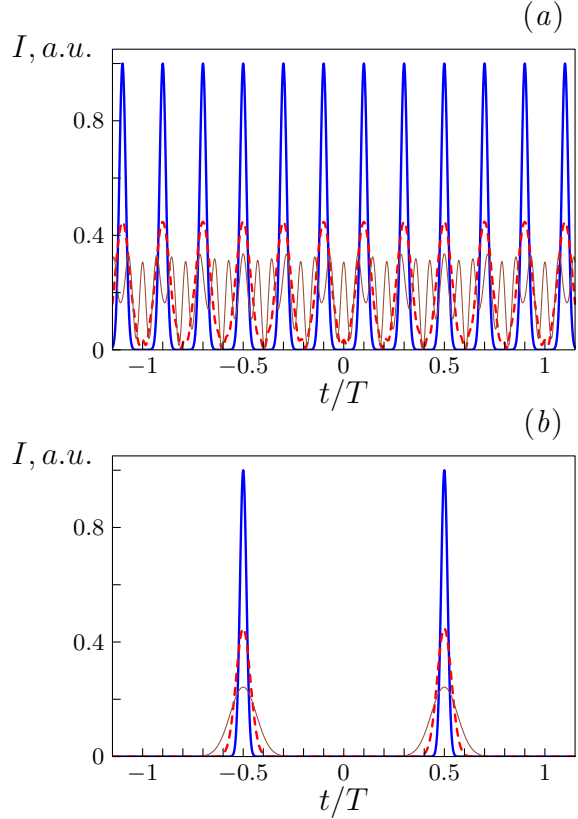}}
\caption{Time dependence of the laser field intensity in arbitrary units
for $n_{0}=10$. \emph{a}) $\alpha=\alpha_{1}$ (thick solid line),
$\alpha=\alpha_{1}+\Delta\alpha$ (dashed line),
$\alpha=\alpha_{1}+2\Delta\alpha$ (thin solid line);
\emph{b}) $\alpha=\pi$
(thick solid line), $\alpha=\pi+\Delta\alpha$ (dashed line),
$\alpha=\pi+2\Delta\alpha$
(thin solid line)
\label{fig-3}}
\end{center}
\end{figure}

\Fref{fig-3} shows the time dependence of the FSF laser
radiation intensity for $\alpha$ close to $\alpha_{1}$ and
$\alpha_{\tilde{m}}=\pi$. As one can see from
\fref{fig-3}\emph{a}, if $\alpha=\alpha_{1}$, the laser radiation
field is a sequence of light pulses that follow each other
with a period of $T/5$, where $T=2\pi/\varpi$. Variation of
$\alpha$ on $\Delta\alpha$ leads to more than twofold
decrease of the amplitude and to the considerable
broadening of the pulses. Another change of $\alpha$
on $\Delta\alpha$ leads to the distortion of the periodical
light pulse  sequence.
In the case of $\alpha=\pi$ (\fref{fig-3}\emph{b}) the laser field
consists of the sequence of light pulses with the
repetition period $T$. Increasing of $|\alpha-\pi|$ leads to
decreasing of the pulse amplitude and to the broadening of pulses.
Therefore, the optimal conditions for the dark resonance formation
are satisfied when the time dependence of the laser
radiation is the periodical sequence of the light
pulses of the maximal amplitude. Both values of $\alpha$
corresponding to pulses of the maximal amplitudes are the
particular cases of the special lasing regimes in which the
FSF laser emits a train of the short pulses~\cite{Yat04-183}. In
general case, the laser emits such pulses when $\alpha=\pi{n}/m$
where $n$ and $m$ are integer.

\begin{figure}
\begin{center}
\includegraphics{{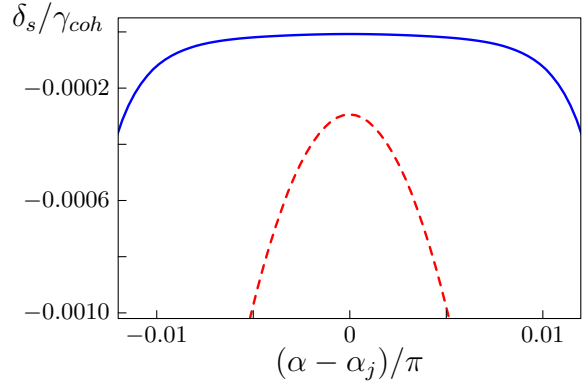}}
\caption{Dependence of the
shift of the dark resonance minimum from the exact two-photon resonance
on $\alpha-\alpha_{j}$ for $j=1$
(solid line) and $j=5$ (dashed line).
Parameters are the same as for \fref{fig-2}\label{fig-4}}
\end{center}
\end{figure}

If $\alpha $ is close to the optimal values $ \alpha_{j}$,
the dependence of the signal on the two-photon detuning is well
described by a Lorentzian curve with a HWHM $\gamma_{coh}$.
Summands with $m\neq\tilde{m}$ lead to the small shift
$\delta_{s}$ of the signal minimum from the two-photon resonance
\begin{equation}
\delta_{s}=- \left[{\frac{\partial }{\partial \delta}\tilde{S}}
\left({\frac{\partial^{2} }{\partial
\delta^{2}}\tilde{S}}\right)^{-1}\right]_{\delta=0} .
\label{eq:shift}
\end{equation}
The dependence of the shift~(\ref{eq:shift}) on the difference
$\alpha-\alpha_{j}$ for parameters corresponding to \fref{fig-2}
is shown in \fref{fig-4}. As one can see,
$\delta_{s}\big|_{\alpha=\pi}$ is approximately 40 times greater
than $\delta_{s}\big|_{\alpha=\pi/\tilde{m}}$. The reason for
this difference   is that at  $\alpha=\pi$ the summands in
the expression (\ref{eq:S_large}) have the same signs due to
multiple of $2\pi$ argument of cosine and sine functions whereas
for $\alpha=\pi/\tilde{m}$ only the part of summands have
the argument which is divisible by $2\pi$. Other summands
has different values of cosine and sine arguments and these values
are distributed approximately homogeneously in the interval
$[0,2\pi]$ if $\tilde{m}$ is large. As a result, the sum of
summands containing Lorentzians with $m\ne{}\tilde{m}$  is
reduced.

\subsection{Broad-band spectrum of the laser radiation}
In this section we suppose that only the condition~(\ref{eq:cr-kv}) 
is valid whereas the relations between
the width of the laser spectrum, the frequency of $|1\rangle
\leftrightarrow |2\rangle$ transition and the optical coherences
relaxation rate $\gamma'$    are arbitrary. In the previous
subsection we established that the main contribution into
the equation \eref{eq:ri-ii-comm-alt-g} and the signal
\eref{eq:Sw} comes from summands with $r_{12,\tilde{m}}$, 
namely from the summands with $m=\tilde{m}$. These summands are
responsible for the two-photon transitions between the
long-lived states in the two-photon resonance conditions when
the frequency difference between spectral components is close to the
frequency of transition $|1\rangle \leftrightarrow |2\rangle$.
The other summands result in the small
shift of the CPT resonance position from the two-photon resonance.
Neglecting this small shift,  from the equations
\eref{eq:ri-ii-comm-alt-g} and \eref{eq:Sw} we obtain
\begin{equation}
\eqalign{
\tilde{S}& = \mathop{\mathrm{Re}}\sum_{n}\left(
\frac{\Omega_{1,n}\Omega_{2,n-\tilde{m}}\rme^{\rmi\varphi_{\tilde{m}}-\rmi\varphi_{n}+\rmi\varphi_{n-\tilde{m}}}}{2\left[\gamma'-\rmi(\delta_{1}-n\varpi)\right]}\right.\\
&\left.\mbox{}+\frac{\Omega_{1,n}\Omega_{2,n-\tilde{m}}\rme^{\rmi\varphi_{\tilde{m}}-\rmi\varphi_{n}+\rmi\varphi_{n-\tilde{m}}}}{2\left[\gamma'+\rmi(\delta_{2}-(n-\tilde{m})\varpi)\right]}\right)R_{12,\tilde{m}}.
}
\label{eq:Swa}
\end{equation}
where
\begin{equation}
\eqalign{
&R_{12,\tilde{m}}=-\frac{1}{4\left[\gamma_{coh}+\gamma_{b}-\rmi{}\left(\delta-\delta_{f}\right)\right]}\\
&\times\sum_{m}\left(\frac{p_{1}\Omega_{1,m}\Omega_{2,m-\tilde{m}}
\rme^{\rmi\varphi_{m}-\rmi\varphi_{\tilde{m}}-\rmi\varphi_{m-\tilde{m}}}}{\gamma'-\rmi[\delta_{1}-m\varpi]}\right.\\
&\left.\mbox{}+ \frac{p_{2}\Omega_{1,m}\Omega_{2,m-\tilde{m}}
\rme^{\rmi\varphi_{m}-\rmi\varphi_{\tilde{m}}-\rmi\varphi_{m-\tilde{m}}}}{\gamma'+\rmi[\delta_{2}-(m-\tilde{m})\varpi]}
\right). } \label{eq:Rii}
\end{equation}
In the expression for $R_{12,\tilde{m}}$, $\gamma_{b}$ is the
field broadening width and $\delta_{f}$ is the dark resonance
light shift. They read
\begin{equation}
\eqalign{
\gamma_{b}=&\frac{\gamma'}{4}\sum\limits_{l}\left(\frac{\Omega_{1,l}^{2}}{\gamma'{}^{2}+(\delta_{1}-l\varpi)^{2}}\right.\\
&\left.+\frac{\Omega_{2,l}^{2}}{\gamma'{}^{2}+(\delta_{2}-l\varpi)^{2}}\right),}
\label{eq:gb}
\end{equation}
\begin{equation}
\eqalign{
\delta_{f}=&\frac{1}{4}\sum\limits_{l}\left(\frac{\Omega_{1,l}^{2}(\delta_{1}-l\varpi)}{\gamma'{}^{2}+(\delta_{1}-l\varpi)^{2}}\right.\\
&\left.-\frac{\Omega_{2,l}^{2}\left(\delta_{2}-l\varpi\right)}{\gamma'{}^{2}+(\delta_{2}-l\varpi)^{2}}\right).}
\label{eq:df}
\end{equation}
It is easy to see from
the equations \eref{eq:Swa} and \eref{eq:Rii} that if
$\alpha$ is determined by the expression (\ref{eq:a-opt}),
the contributions with different values of $n$, $m$ into the
signal are summed, and the signal is maximal. If $\alpha$ differs
from the optimal values defined by expression \eref{eq:a-opt}, the
signal decreases due to the presence of the summands with
different phases.

\begin{figure}
\begin{center}
\includegraphics{{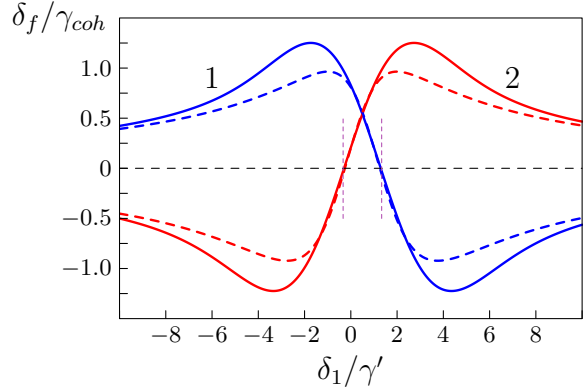}}
\caption{The dependence of the light shift on the detuning
$\delta_{1}$ of the maximal intensity spectral component
frequency from the frequency of transition between states
$|1\rangle$ and $|3\rangle$. Parameters:
$\gamma'=\omega_{21}=1000\gamma_{coh}$, $\varpi=20\gamma_{coh}$,
$n_{0}=400$. The Rabi frequencies corresponding to the
maximal intensity spectral components  are: 1~--
$\Omega_{1,0}=5\gamma_{coh}$, $\Omega_{2,0}=10\gamma_{coh}$, 2~--
$\Omega_{1,0}=10\gamma_{coh}$, $\Omega_{2,0}=5\gamma_{coh}$. Solid
line~-- the expression~(\ref{eq:df}) evaluation, dashed
line~-- the expression~(\ref{eq:shift_lor}) evaluation. The magnitudes
of $\delta_{z}$ calculated using~(\ref{eq:dz}) are shown by
vertical segments. \label{fig-5}}
\end{center}
\end{figure}

If $\gamma'\gg{}\varpi$ and $n_{0}\gg1$, the summation in
the expression~(\ref{eq:df}) for the light shift can be
replaced by the integration over $l$. The analytical expression of
the integral for the Gaussian spectral distribution of
intensity cannot be obtained. For estimations we
replace the Gaussian spectral distribution by the
Lorentzian distribution
\begin{equation}
\Omega_{j,l}^{2}={\Omega_{j,0}^{2}}\left(1+\frac{2\pi{}l^{2}}{n_{0}^{2}}\right)^{-1}.
\label{eq:Lor}
\end{equation}
Replacing summation by integration in the expression
\eref{eq:df} and taking into account the expression
\eref{eq:Lor} we find
\begin{equation}
\delta_{f}=\Delta_{1}-\Delta_{2},
\label{eq:shift_lor}
\end{equation}
where
\begin{equation}
\Delta_{j}=\frac{\Omega_{j,0}^{2}n_{0}\delta_{j}\sqrt{\pi^{3}}\left[\left(\sqrt{2\pi\sqrt{2}}\gamma'-n_{0}\varpi2^{1/4}\right)^{2}+
2\pi\sqrt{2}\delta_{j}^{2}\right]}{4\left[4\pi^{2}\left(\delta_{j}^{2}
+\gamma'^{2}\right)^{2}+n_{0}^{2}\varpi^{2}\left(n_{0}^{2}\varpi^{2}
+4\pi\delta_{j}^{2}-4\pi\gamma'^{2}\right)\right]}.
\label{eq:Delta}
\end{equation}
\Fref{fig-5} shows examples of the light shift dependencies
on the detuning $\delta_{1}$ of the laser radiation
spectrum maximum from the frequency of transition $|1\rangle
\leftrightarrow |3\rangle$ calculated using the expressions
\eref{eq:df} and \eref{eq:shift_lor}.
One can see that the results calculated using both formulae
are in good correspondence when $\delta_{1}$ is small (of order of
$\gamma'$). This enables us to find from
\eref{eq:shift_lor} the value of $\delta_{1}=\delta_{z}$ for which
the light shifts caused by the field acting at
$|1\rangle\leftrightarrow|3\rangle$ and
$|2\rangle\leftrightarrow|3\rangle$ transitions compensate
one another. If the spectrum is broad-band
($n_{0}\varpi\gg\gamma'$) and if the dipole matrix elements of the
transition $|1\rangle \leftrightarrow |3\rangle$, $|2\rangle
\leftrightarrow |3\rangle$ are different, we find
\begin{equation}
\delta_{z}=\frac{\omega_{21}\Omega_{2,0}^{2}}{\Omega_{2,0}^{2}-\Omega_{1,0}^{2}}=
\frac{\omega_{21}d_{23}^{2}}{d_{23}^{2}-d_{13}^{2}}.
\label{eq:dz}
\end{equation}
Here we used the equality
$\Omega_{1,0}/\Omega_{2,0}=d_{13}/d_{23}$. Let us note that if
$\delta_{1}=\delta_{z}$ and the parameters of the atom-field
interaction correspond to \fref{fig-5}, the magnitude of the
signal is close to a maximum.

\section{Small optical coherence relaxation rate}
\label{s:small} 
\begin{figure}
\begin{center}
\includegraphics{{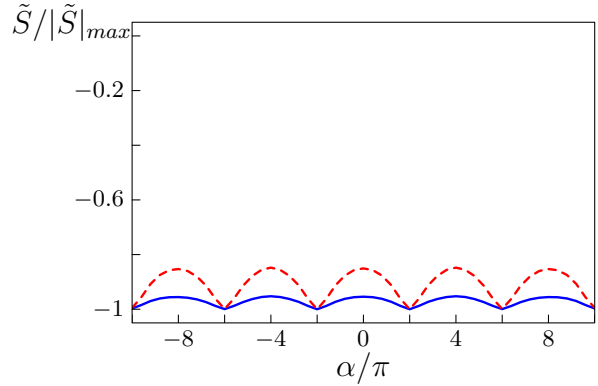}}
\caption{The dependence of the normalized signal at
two-photon resonance ($\delta=0$) on $\alpha$ for $n_{0}=5$,
$\tilde{m}=5$, $\varpi=200\gamma_{coh}$, $p_{1}=p_{2}=0.5$,
$kv_{0}=5000\gamma_{coh}$, $\nu=0$. Solid line corresponds to
$\gamma'=10\gamma_{coh}$, dashed line corresponds to
$\gamma'=20\gamma_{coh}$ \label{fig-6}}
\end{center}
\end{figure}

The case when $\gamma'$ is small in comparison
with the Doppler width of the spectrum,
\begin{equation}
\gamma'\leq{}kv_{0},\qquad
\label{eq:crit-kvg}
\end{equation}
we consider in the approximation \eref{eq:cr-broad}
neglecting the field broadening and the light shifts. From
the expressions \eref{eq:ri-ii-comm-alt-g} and \eref{eq:Sw}
we find that
\begin{equation}
\eqalign{
\fl\tilde{S}=&-\frac{1}{8}\mathop{\mathrm{Re}}\sum_{n,m,l}\int\limits_{}^{}
\Omega_{1,l}\Omega_{2,l-m}\Omega_{1,n}\Omega_{2,n-m}
\rme^{\rmi\varphi_{l}-\rmi\varphi_{l-m}-\rmi\varphi_{n}+\rmi\varphi_{n-m}}\\
\fl&\times\left(2\gamma'-\rmi\delta-\rmi\tilde{m}\varpi+\rmi{}m\varpi\right)
\left[\gamma'-\rmi\left(\delta_{1}-n\varpi+kv\right)\right]^{-1}\\
\fl&\times\left[\gamma'+\rmi\left(\delta_{1}-\delta-n\varpi+m\varpi
-\tilde{m}\varpi+kv\right)\right]^{-1} \\
\fl&\mbox{}\times\left(\frac{p_{2}}{\gamma'+
\rmi\left[\delta_{1}-\delta-(l+\tilde{m}-m)\varpi+kv\right]}\right.\\
\fl&\left.\mbox{}+\frac{\nu{}p_{2}F(\left[\delta_{1}-\delta-l\varpi+m\varpi-
\tilde{m}\varpi\right]/\gamma')}{\gamma'\left[\gamma_{coh}
-\rmi\left(\delta+\tilde{m}\varpi-m\varpi\right)\right]}
+\mbox{}\right.\\
\fl&\left.\mbox{}+\frac{p_{1}}{\gamma'
-\rmi\left[\delta_{1}-l\varpi+kv\right]}
+\frac{\nu{}p_{1}F(\left[\delta_{1}-l\varpi\right]/\gamma')^{*}}{\gamma'\left[\gamma_{coh}
-\rmi\left(\delta+\tilde{m}\varpi-m\varpi\right)\right]}
\right)\\
\fl&\times\frac{M(v)}{\gamma_{coh}
+\nu-\rmi\left(\delta+\tilde{m}\varpi-m\varpi\right)}\,dv. }
\label{eq:Spoly}
\end{equation}
Here we introduce the function of dimensionless variable~$x$
\begin{equation}
F(x)=\int\limits_{-\infty}^{\infty}\frac{\gamma'M(v)}{\gamma'+
\rmi(\gamma'x+kv)}\,dv.
\label{eq:F}
\end{equation}
If $\gamma'\ll{}kv_{0}$ and $\gamma'\ll{}\varpi$ but
$\gamma'\gg{}\gamma_{coh}$, the main contribution into the
signal~(\ref{eq:Spoly}) comes from the summands with
$m=\tilde{m}$, $l=n$. Therefore the triple summation is reduced to
a single one and  the result of summing does not depend on
the phases of the spectral components of the field. \Fref{fig-6}
shows the weak, more weak than in \fref{fig-2}, dependence of the
signal at two-photon resonance on $\alpha$ that decreases with
decreasing of $\gamma'$.

\begin{figure}
\begin{center}
\includegraphics{{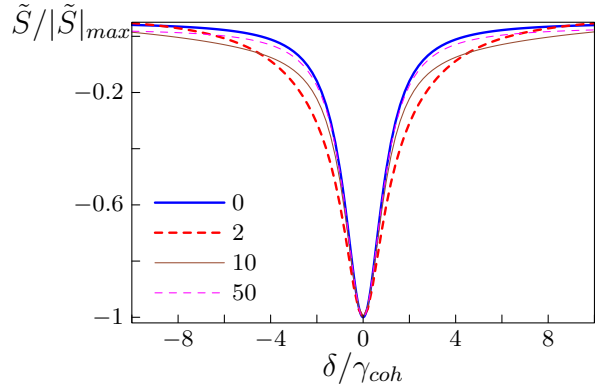}}
\caption{The dependence of the signal (normalized to unit
magnitude at two-photon resonance) on the two-photon detuning for
$n_{0}=10$, $\tilde{m}=10$, $\varpi=50\gamma_{coh}$,
$p_{1}=p_{2}=0.5$, $kv_{0}=2000\gamma_{coh}$, $\alpha=\pi$,
$\gamma'=200\gamma_{coh}$. The curves are marked by the magnitude
of $\nu/\gamma_{coh}$\label{fig-7}}
\end{center}
\end{figure}

\begin{figure}
\begin{center}
\includegraphics{{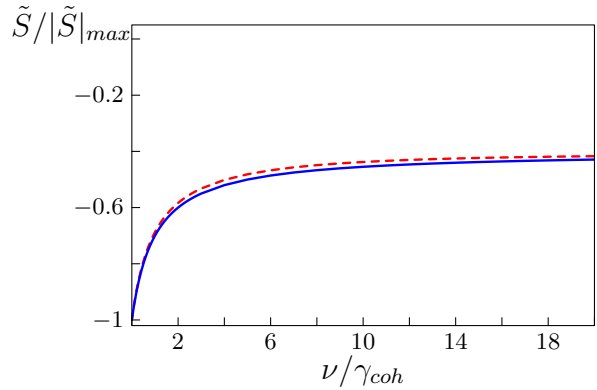}}
\caption{The dependence of the normalized signal on
the collision preserving frequency at two-photon resonance
condition ($\delta=0$) for  $n_{0}=10$, $\tilde{m}=10$,
$\varpi=50\gamma_{coh}$, $p_{1}=p_{2}=0.5$,
$kv_{0}=2000\gamma_{coh}$, $\gamma'=200\gamma_{coh}$; solid line
corresponds to $\alpha=\pi/\tilde{m}$, dashed line corresponds to
$\alpha=\pi$ \label{fig-8}}
\end{center}
\end{figure}

In contrast to the case of large optical coherence
relaxation rate $\gamma'\gg{}kv_{0}$ considered above, when the
signal does not depend on the rate $\nu$ of coherence preserving
collisions, in the case   $\gamma'\le{}kv_{0}$ the signal strongly
depends on $\nu$. Examples of dependencies of the signal on
$\delta$ for $\gamma'>\varpi$ are shown in \fref{fig-7}. One can
see that increasing of $\nu$ leads to the initial increasing and
subsequent decreasing of the dark resonance HWHM.
\Fref{fig-8} shows the dependence of the signal on the collision
frequency if the condition of two-photon resonance is fulfilled.
One can see that increasing of $\nu$ leads to decreasing
of registered signal, fast initially and slow at
$\nu\gg\gamma_{coh}$.

This effect can be explained qualitatively in the following
way. If $\gamma'\gg k v$, the certain atom after each collision
remains being interacting with the same components of laser
radiation as before the collision, i.e., with all the components
whose Doppler-shifted frequencies $\omega_n-k v$ are close enough
to the optical transition frequencies $\omega_{31}$ or
$\omega_{32}$ in the atom, $|\omega_n-\omega_{31,32}-k
v|<<\gamma'$, therefore velocity changes due to   collisions
with the buffer gas atoms do not lead to any additional
broadening or shift of the dark resonances.  If $\gamma' \leq k
v$, any collision will modify the set of the laser
components that interacts with the atom.
Therefore the collisional 
rate $\nu$ occurs to be an important parameter determining
the dark resonance properties.

We should keep in mind that parameters $\gamma'$, $\gamma_{coh}$
and $\nu$ are not independent. While $\gamma'$ and $\nu$
increase linearly with the buffer gas pressure, $\gamma_{coh}$
decreases approximately as a square root of the pressure. Without
buffer gas, $\gamma_{coh}$ is approximately equal to the inverse
time of flight of an atom through the laser beam, and in buffered
cells it is approximately equal to the inverse time of diffusion
of an atom out of the beam (this dependence occurs to be more
complicated if an atom can return back to the laser beam without
loss of the coherence, see \cite{Xiao,Rom}). When we consider the
coherence preserving collisions rate as an independent
parameter calculating the dependencies shown in \fref{fig-8}
we just estimate its influence on the registered signal magnitude.

\section{Conclusions}
\label{s:conclusions}

We present the    theory      of  the dark  resonances in a
fluorescence of a three-level atom gas interacting with a
polychromatic field of a frequency shifted feedback (FSF) laser.
We show that conditions for the resonance observation are
optimal when the phase relations between the laser spectral
components provide generation of a light pulses train.
We study analytically the field broadening and the light
shift of the resonances.

The expressions for the field broadening width and
light shift of the dark resonances are found for the large
(in comparison with Doppler width) optical coherence
relaxation rate.

We found the condition on the detuning of the spectral component with maximal
intensity from the transition frequency between one of the long-lived states and
the short-lived state that provides zero value of the light shift.
This condition does not depend on the laser field strength.

It is shown that increasing of the coherence preserving collisions
rate leads to decreasing of the signal. This decreasing is quick
at small collision rate and becomes more slow with
increasing of the collision rate.

\ack

The work was financially supported by State Foundation of Fundamental Researches
of Ukraine (project F28.2/035), Russian Foundation for Basic Researches
(project 09-02-90465\_Ukr\_f\_a), FTsP ``Scientific and Pedagogical personnel
of innovative Russia'' (project P2326 from 16.11.2009), Russian President Grant for
Young Candidates of Sciences (project MK-5318.2010.2).


\section*{References}
\begin{thebibliography}{10}

\bibitem{Ari76-333}Arimondo E and Orriols G 1976 \emph{Nuovo Cimento Lett.} \textbf{17}
333--338

\bibitem{Alz76-5}Alzetta G, Gozzini A, Moi L and Orriols G 1976
\NC B Ser. 2 \textbf{36} 5--20

\bibitem{Gra78-218} Gray H R, Whitley R W and Stroud C R Jr 1978
\emph{Optics Lett.} \textbf{3} 218--220

\bibitem{Orr79-5} Orriols G 1979 \NC B Ser. 2 \textbf{53} 1--24

\bibitem{Alz79-209} Alzetta G, Moi L and Orriols G 1979 \NC B Ser. 2 \textbf{52} 209--218

\bibitem{Koc86} Kocharovskaya O A and Khanin  Ya I 1986 \emph{Sov. Phys. JETP}
\textbf{63} 945--950


\bibitem{Gor89}
Gorny\u{\i} M B, Matisov B G and Rozhdestvenski\u{\i}  Yu V 1989
\emph{Sov. Phys. JETP} \textbf{68} 728--732

\bibitem{Bol91-2593} Boller K-J, Imamoglu A and Harris S E 1991
\PRL \textbf{66} 2593--2596

\bibitem{Har97-36} Harris S E 1997 \emph{Phys. Today} \textbf{50} 36--42

\bibitem{Ore84-690} Oreg J, Hioe F T and Eberly  J H 1984 \PR A \textbf{29} 690--697


\bibitem{Gau88-463} Gaubatz U, Rudecki P, Becker M, Schiemann S, Kulz M and Bergmann K 1988 \emph{Chem. Phys. Lett.} \textbf{149} 463--468

\bibitem{Ber98-1003} Bergmann K, Theur H and Shore B W 1998 \RMP \textbf{70} 1003--1025


\bibitem{Vit01} Vitanov N V, Halfmann T, Shore B W, Bergmann K 2001
\emph{Annu. Rev. Phys. Chem.} \textbf{52} 763--809

\bibitem{Asp88-826} Aspect A, Arimondo E, Kaiser R, Vansteenkiste N and Cohen-Tannoudji C 1988 \PRL \textbf{61} 826--829

\bibitem{Kna01-1545}Knappe S, Wynands R, Kitching J, Robinson H G and Hollberg L 2001
\JOSA B \textbf{18} 1545--1553

\bibitem{Nag98-31} Nagel A, Graf L, Naumov A, Mariotti E, Biancalana V, Meschede D and Wynands R 1998
\emph{Europhys. Lett.} \textbf{44} 31--36

\bibitem{Sch04-6409} Schwindt P D D, Knappe S, Shah V,
Hollberg L, Kitching J, Liew L-A and Moreland J 2004
\emph{Appl. Phys. Lett.} \textbf{85} 6409--6411


\bibitem{Dic53} Dicke R H 1953 \PR \textbf{89} 472--473


\bibitem{Erh01-043813} Erhard M and Helm H 2001 \PR A \textbf{63} 043814

\bibitem{Mer03} Merimaa M, Lindvall T, Tittonen I and Ikonen E 2003
\JOSA B \textbf{20} 273--279

\bibitem{Alz04} Alzetta G, Gozzini S, Lucchesini A,
Cartaleva S, Karaulanov T, Marinelli C and Moi L 2004
\PR A \textbf{69}, 063815

\bibitem{Sau} Sautenkov V A, Rostovtsev Yu V, Ye C Y, Welch G R,
Kocharovskaya O and Scully M O 2005 \PR A \textbf{71} 063804

\bibitem{Ari06-169} Arissian L and Diels J-C 2006 \emph{Optics Communications} \textbf{264} 169--173

\bibitem{Vla06-609} Vladimirova  Yu V, Grishanin B A, Zadkov V N,
Biancalana V, Bevilacqua  G, Dancheva  Y and Moi L 2006 \emph{JETP} \textbf{103} 528--538

\bibitem{Auz}
Auzinsh M, Malitskiy  R A, Matsnev I V, Negriyko  A M, Romanenko V I and
Yatsenko L P 2009 \emph{Ukr. J. Phys.}  \textbf{54} 974--982


\bibitem{Yat04-183} Yatsenko L P, Shore B W and Bergmann K 2004
\emph{Optics Communications} \textbf{236} 183--202

\bibitem{Rau66-1176} Rautian S G 1967 Sov. Phys. \emph{JETP} \textbf{24}
788--796

\bibitem{Rau66-209} Rautian S G and Sobel'man I I 1967 \emph{Sov. Phys. Usp.} \textbf{9}
701--716

\bibitem{Xiao} Xiao Y, Novikova I, Phillips D F, Walsworth R L
2006 \PRL \textbf{96} 043601

\bibitem{Rom}
Romanenko V I, Romanenko A V, Yatsenko L P 2010 \emph{Ukr. J. Phys.}
\textbf{55} 393--402



\end{thebibliography}
\end{document}